\documentclass[%
 reprint,
 amsmath,amssymb,
 aps,
]{revtex4-1}

\usepackage{graphicx}
\usepackage{dcolumn}
\usepackage{bm}
\usepackage[tight,footnotesize]{subfigure}

\begin{document}


\title{Influence of Laser Intensity Fluctuation on Single-Cesium Atom Trapping Lifetime in a 1064-nm Microscopic Optical Tweezer}

\author{Rui Sun,${}^{1}$ Xin Wang,${}^{1}$ Kong Zhang,${}^{1}$ Jun He,${}^{1,2}$}

\author{Junmin Wang${}^{1,2}$}
 \email{wwjjmm@sxu.edu.cn}
\affiliation{%
1 State Key Laboratory of Quantum Optics and Quantum Optics Devices, and Institute of Opto-Electronics, Shanxi University, Taiyuan 030006, China
}%

\affiliation{%
2 Collaborative Innovation Center of Extreme Optics of the Ministry of Education and Shanxi Province, Shanxi University, Taiyuan 030006, China
}%

\begin{abstract}
An optical tweezer composed of a strongly focused single-spatial-mode Gaussian beam of a red-detuned 1064-nm laser can confine a single-cesium (Cs) atom at the strongest point of the light intensity. We can use this for coherent manipulation of single-quantum bits and single-photon sources. The trapping lifetime of the atoms in the optical tweezers is very short due to the impact of the background atoms, the laser intensity fluctuation of optical tweezer and the residual thermal motion of the atoms. In this paper, we analyzed the influence of the background pressure, the trap frequency of optical tweezers and the parametric heating of the optical tweezer on the atomic trapping lifetime. Combined with the external feedback loop based on an acousto-optical modulator (AOM), the intensity fluctuation of the 1064-nm laser in the time domain was suppressed from $\pm$ 3.360$\%$ to $\pm$ 0.064$\%$, and the suppression bandwidth reached approximately 33 kHz. The trapping lifetime of a single Cs atom in the microscopic optical tweezer was extended from 4.04 s to 6.34 s.

\end{abstract}

\maketitle


\section{\label{sec:level1}Introduction}

The physical implementation of a single-photon source has important application value in the fundamental research of quantum optics and linear quantum computation, especially the controllable triggered single-photon source as its core quantum source. Generally, one can generate single-photons via single atom, single molecule, single ion, single quantum dots or parametric down-conversion. Compared with the said ways, single atoms have advantages of narrowband, matching atom transition lines and weak coupling of neutral ground-state atoms with background light and external electromagnetic fields. single atom source based on the captured single atom in an optical tweezer\cite{1,2,3} paves the way for quantum repeaters, quantum teleportation, quantum secure communications and linear quantum computing. In 1975, Hansch and Schawlow first put forward the use of the laser to cool neutral atoms \cite{4}. In 1987, the research group led by Steven Chu cooled and captured neutral Sodium atoms with magnetic optical trap (MOT) for the first time \cite{5}. In 1994, Kimble’s team first achieved the cooling and capture of a single atom \cite{6}. In 2016, Browaeys group constructed a two-dimensional atom array composed of 50 single atoms trapped in optical tweezers \cite{7}. In 2018, Browaeys group constructed a three-dimensional atom array composed of 72 single atoms trapped in optical tweezers, optionally manipulating the spatial position of each atom \cite{8}. In our system, we have already captured \cite{2,9} and transferred \cite{10,11} single atoms to optical tweezer efficiently, built cesium (Cs) magic-wavelength optical dipole trap \cite{2,3}, and finally achieved triggered single-photon source at 852 nm based on single atom manipulation \cite{3,12}.

When manipulating the single atoms in the optical tweezers, it is required that the atoms be captured we must capture all the atoms in the tweezers before finishing all the operations. Therefore, it is important to prolong the trapping lifetime of atoms in the optical tweezer, to improve the experimental efficiency and the experimental accuracy.

The trapping lifetime of atom in the optical tweezer is limited by lots of factors such as recoil heating, laser intensity fluctuation, the vacuum degree of atomic vapor cell, laser beam pointing stability and the residual atom thermal motion. One can suppress the laser intensity fluctuation \cite{13}, improve  background vacuum degree, improve the laser pointing stability, increase the laser power and adopt the polarization gradient cooling \cite{14} to prolong the trapping lifetime of a single atom in the optical tweezer. In this work, we discussed the influence of the vacuum degree and the light intensity fluctuation of 1064-nm optical tweezer on atom trapping lifetime and the experimental methods of improving the fluctuation of the laser intensity.

Generally, one can suppress the laser intensity fluctuation with optical mode cleaner \cite{15}, optical injection locking \cite{16} or acousto-optical modulator (AOM) feedback \cite{13,17}. Thanks to its flexibility and simplified experimental setup, the AOM feedback becomes the best candidate to suppress the laser intensity fluctuation only by controlling the diffraction efficiency of AOM.

\section{\label{sec:level1}single atom Magnetic Optical Trap}

 Trapping lifetime of the ground state atoms in the optical tweezers is limited due to the influence of background atoms collision, the parametric heating of the optical tweezer and the residual thermal motion of the atoms.

The individual atom trapped in the optical tweezer will collide with the background atoms in the Cs gas cell. In this process the single atom gains kinetic energy, heats up and eventually escapes from the optical tweezer, resulting in the reduction of the atomic trapping lifetime. In the experiment, we can reduce the collision probability of induced heating by improving the background vacuum degree of the cesium atom gas cell and reducing the number of background atoms. This can extend the trapping lifetime to some extent.

Since the probe of the thermal ionization gauge in the vacuum system is far away from the cesium atom gas cell, the measured vacuum degree may be different from the actual vacuum degree in the Cs atomic vapor cell. By measuring the typical time of the magneto-optical trap (MOT), we can know the pressure level in the Cs atomic gas cell more accurately. In our scheme, due to MOT capture the atoms in the cell, it is impossible to isolate trapped atoms from background atoms. Thus, we measure and fit the loading curve instead of using release $\&$ recapture to get the typical time of MOT \cite{18}. The background pressure is related to the atomic density N in the vapor cell \cite{18}
\begin{equation}
N=\frac{1}{\tau\sigma}\sqrt\frac{M}{3k_BT}
\end{equation}
where $\tau$ is the atomic trapping lifetime in the MOT, $\sigma$ is the atomic collision cross-section of Cs, M is the mass of Cs atom, $k_B$ is the Boltzmann constant and T is the romm temperature, according to the pressure formula $p=Nk_BT$ the pressure (p) in the vapor cell is \cite{18}
\begin{equation}
p=\frac{\sqrt{3Mk_BT}}{3\tau\sigma}
\end{equation}

\begin{figure}
	\includegraphics[width=.4\textwidth]{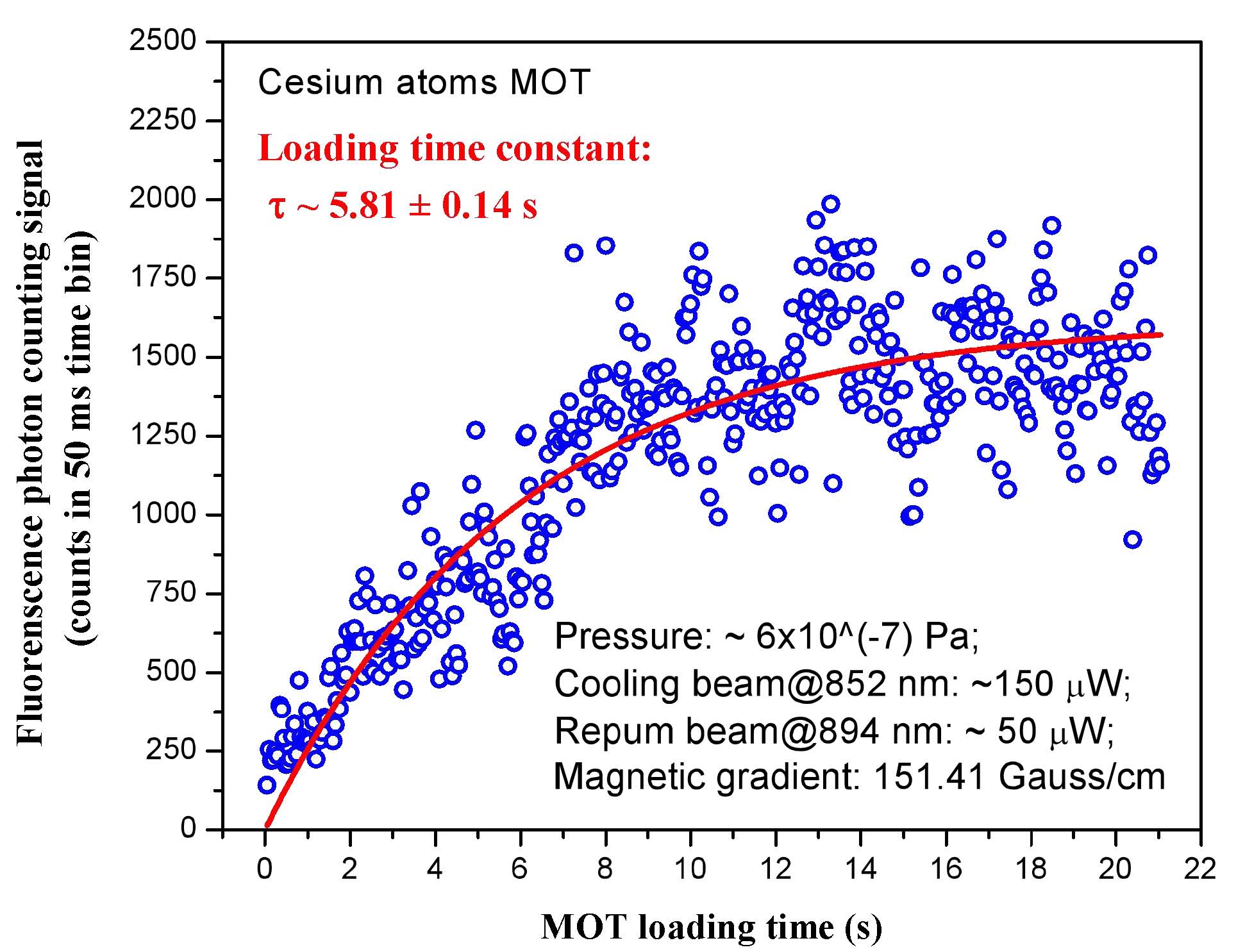}
	\caption{Loading curve of the magneto-optical trap (MOT). The MOT is empty when the quadrupole magnetic field is turned off. After the quadrupole magnetic field is turned on, atoms are gradually loaded into the MOT. The typical time constant of the MOT is 5.81 $\pm$ 0.14 s, and the corresponding pressure in the Cs vapor cell is approximately 6 $\times$ $10^{−7}$ Pa.}
	\label{figure1}
\end{figure}

Figure~\ref{figure1} shows the loading curve of the magneto-optical trap when the cooling laser power of the MOT is 150 $\mu$ W, and the magnetic field gradient is 151.41 Gauss/cm. For atom number $N'(t=0)=0$ as initial condition and fitting the curve with $N'(t)=N_S(1-e^{-t/\tau})$ \cite{18}, for $N_S$ is the steady-state number. The typical time of the MOT is 5.81 $\pm$ 0.14 s, and the pressure in the cell is about 6 $\times$ $10^{-7}$ Pa. The temperature of atoms in the MOT is $\sim$105 $\mu$ K under similar conditions \cite{19}.

\begin{figure*}[!htb]
        \centering
        \subfigure[] {
                \includegraphics[width=0.8\columnwidth]{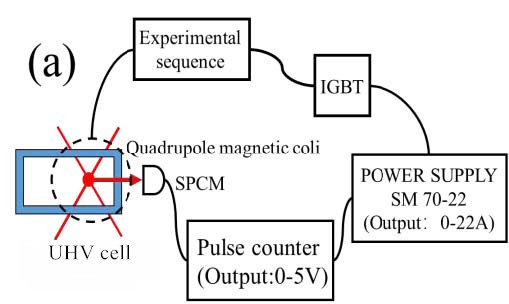}
                \label{fig:overview2_3}
            }
        \subfigure[]  {
                \includegraphics[width=0.8\columnwidth]{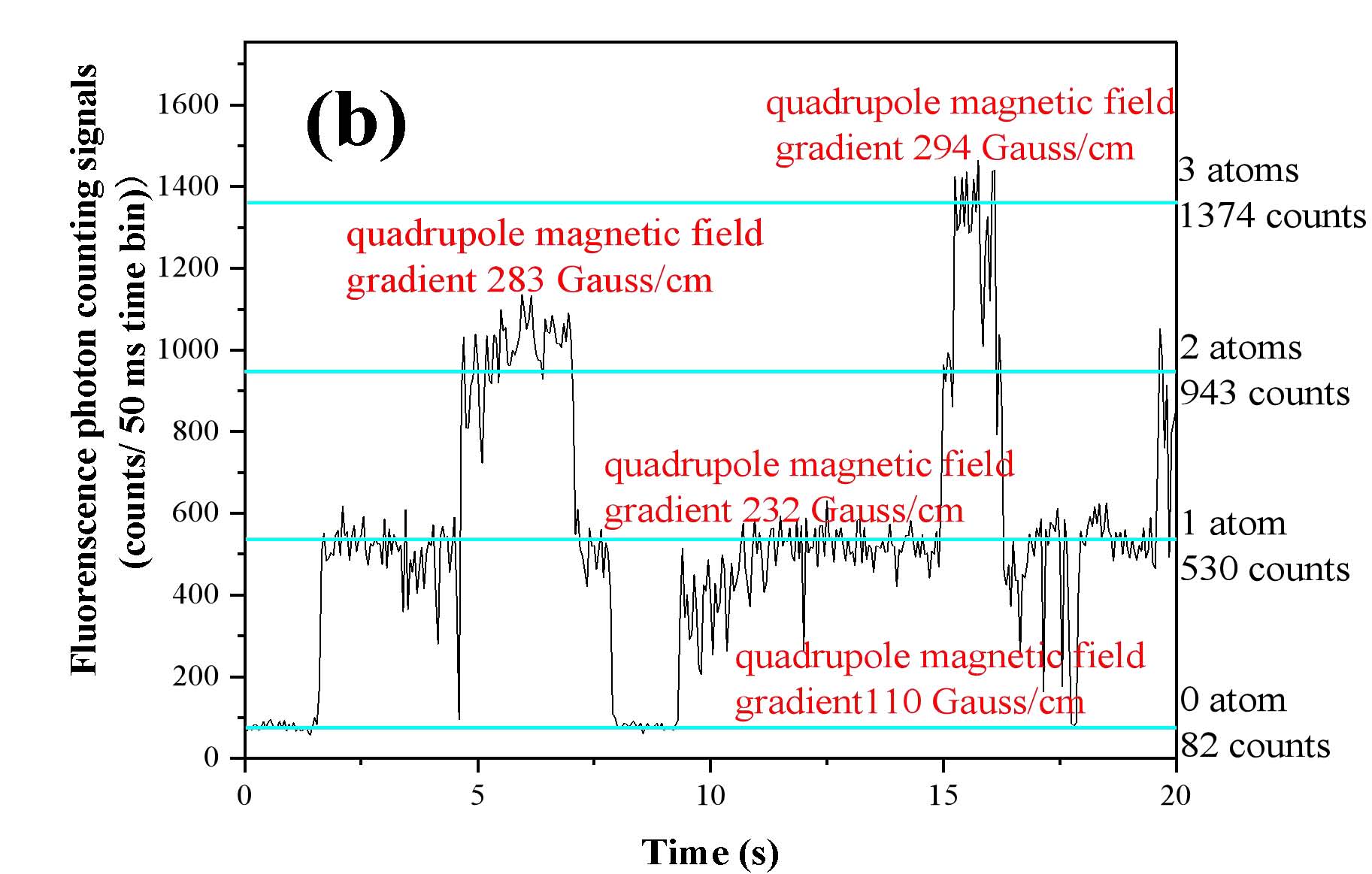}
                \label{fig:overview2_4}
        }
        \vspace{-0.15in}
        \caption{single atom loading trigger loop. (a) Pulse counter adjusts the output voltage according to different photon counts. We use the output analog voltage to control the output current of power supply and the quadrupole magnetic field gradient. Keys to (a) single-photon counting module (SPCM); insulated gate bipolar transistor (IGBT); (b) The trigger loop dynamically changes the loading rate of atoms in the MOT by adjusting the quadrupole magnetic field gradient.}
        \label{figure2}
\end{figure*}

\begin{table*}[]
\begin{tabular}{|c|c|c|c|c|c|c|c|c|}
\hline
\begin{tabular}[c]{@{}c@{}}Atom \\ Numbers\end{tabular} & \multicolumn{2}{c|}{\begin{tabular}[c]{@{}c@{}}Photon Counts\\ (Counts/50ms)\end{tabular}}                     & \multicolumn{2}{c|}{\begin{tabular}[c]{@{}c@{}}Pulse Counter \\ Analog Output\\ Voltage (V)\end{tabular}}      & \multicolumn{2}{c|}{\begin{tabular}[c]{@{}c@{}}Power Supply\\ Output Current \\ (A)\end{tabular}}             & \multicolumn{2}{c|}{\begin{tabular}[c]{@{}c@{}}Quadrupole Magnetic \\ Field Gradient \\ (Gauss/cm)\end{tabular}} \\ \hline
-                                                       & \begin{tabular}[c]{@{}c@{}}Lower \\ Bound\end{tabular} & \begin{tabular}[c]{@{}c@{}}Upper\\ Bound\end{tabular} & \begin{tabular}[c]{@{}c@{}}Lower \\ Bound\end{tabular} & \begin{tabular}[c]{@{}c@{}}Upper\\ Bound\end{tabular} & \begin{tabular}[c]{@{}c@{}}Lower\\ Bound\end{tabular} & \begin{tabular}[c]{@{}c@{}}Upper\\ Bound\end{tabular} & \begin{tabular}[c]{@{}c@{}}Lower\\ Bound\end{tabular}   & \begin{tabular}[c]{@{}c@{}}Upper\\ Bound\end{tabular}  \\ \hline
0                                                       & 0                                                      & 400                                                   & 1.50                                                   & 1.79                                                  & 7.0                                                   & 7.9                                                   & 103                                                     & 116                                                    \\ \hline
1                                                       & 500                                                    & 600                                                   & 3.80                                                   & 3.80                                                  & 17.3                                                  & 17.3                                                  & 232                                                     & 232                                                    \\ \hline
2                                                       & 601                                                    & 1500                                                  & 4.00                                                   & 4.50                                                  & 18.6                                                  & 20.7                                                  & 273                                                     & 304                                                    \\ \hline
3                                                       & 1501                                                   & 2500                                                  & 4.51                                                   & 4.70                                                  & 20.8                                                  & 21.5                                                  & 306                                                     & 316                                                    \\ \hline
\end{tabular}
\caption{Pulse counter and power supply parameter setting.}
\label{table:table1}
\end{table*}

\begin{figure*}[!htb]
        \centering
        \subfigure {
                \includegraphics[width=0.75\columnwidth]{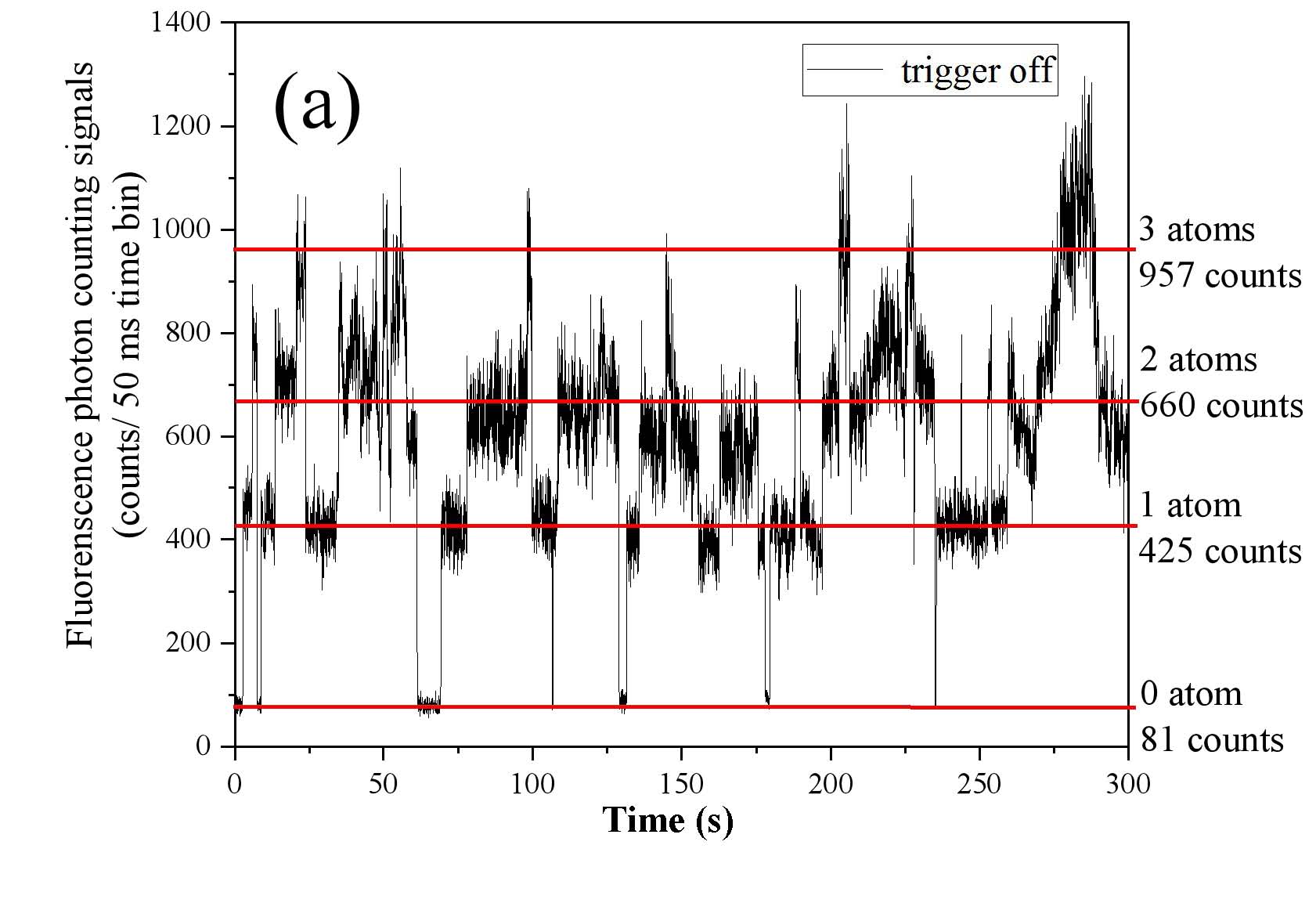}
                \label{fig:overview2_3}
            }
        \subfigure  {
                \includegraphics[width=0.63\columnwidth]{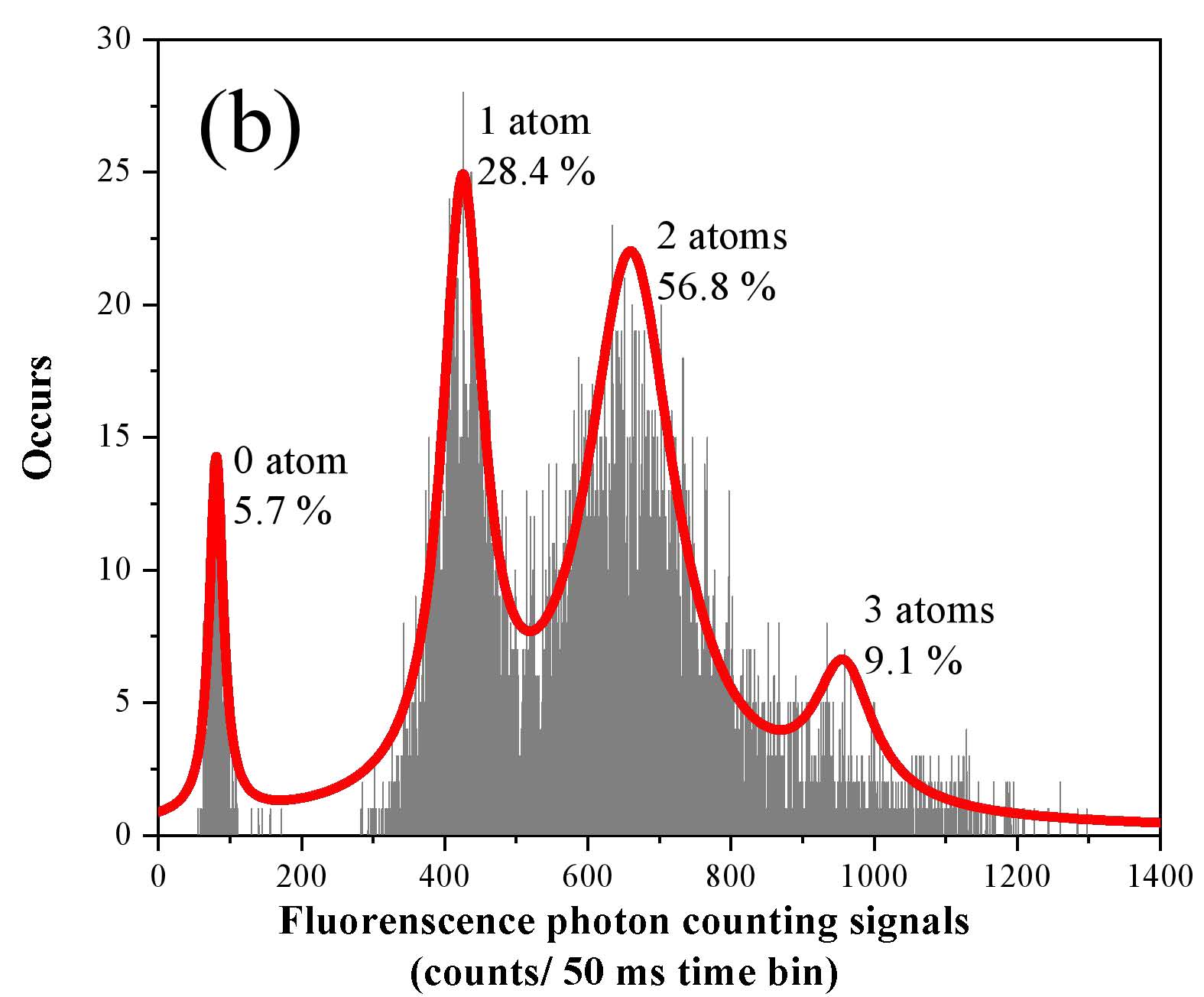}
                \label{fig:overview2_4}
        }

        \subfigure {
                \includegraphics[width=0.75\columnwidth]{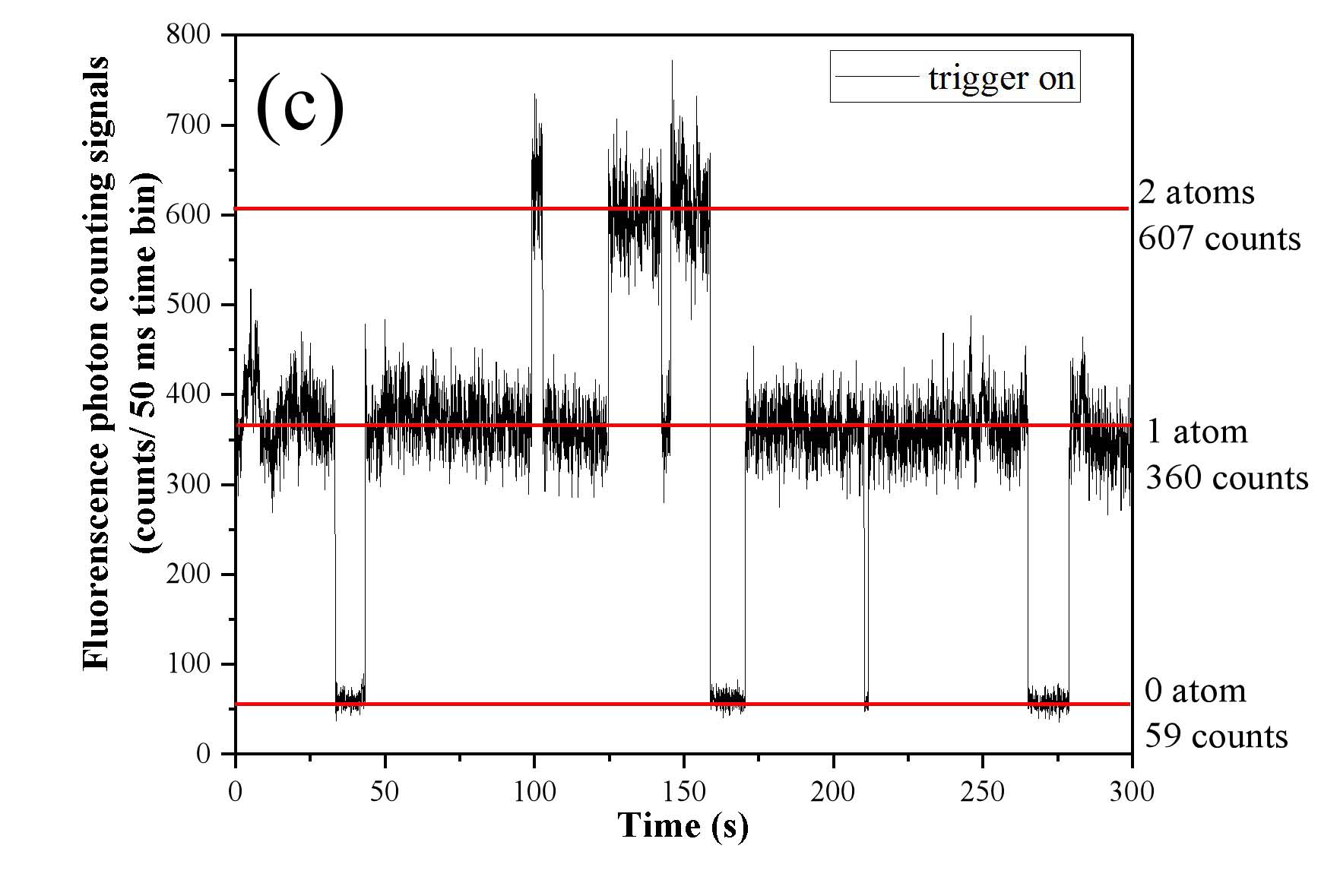}
                \label{fig:overview2_4}
        }
        \subfigure {
                \includegraphics[width=0.60\columnwidth]{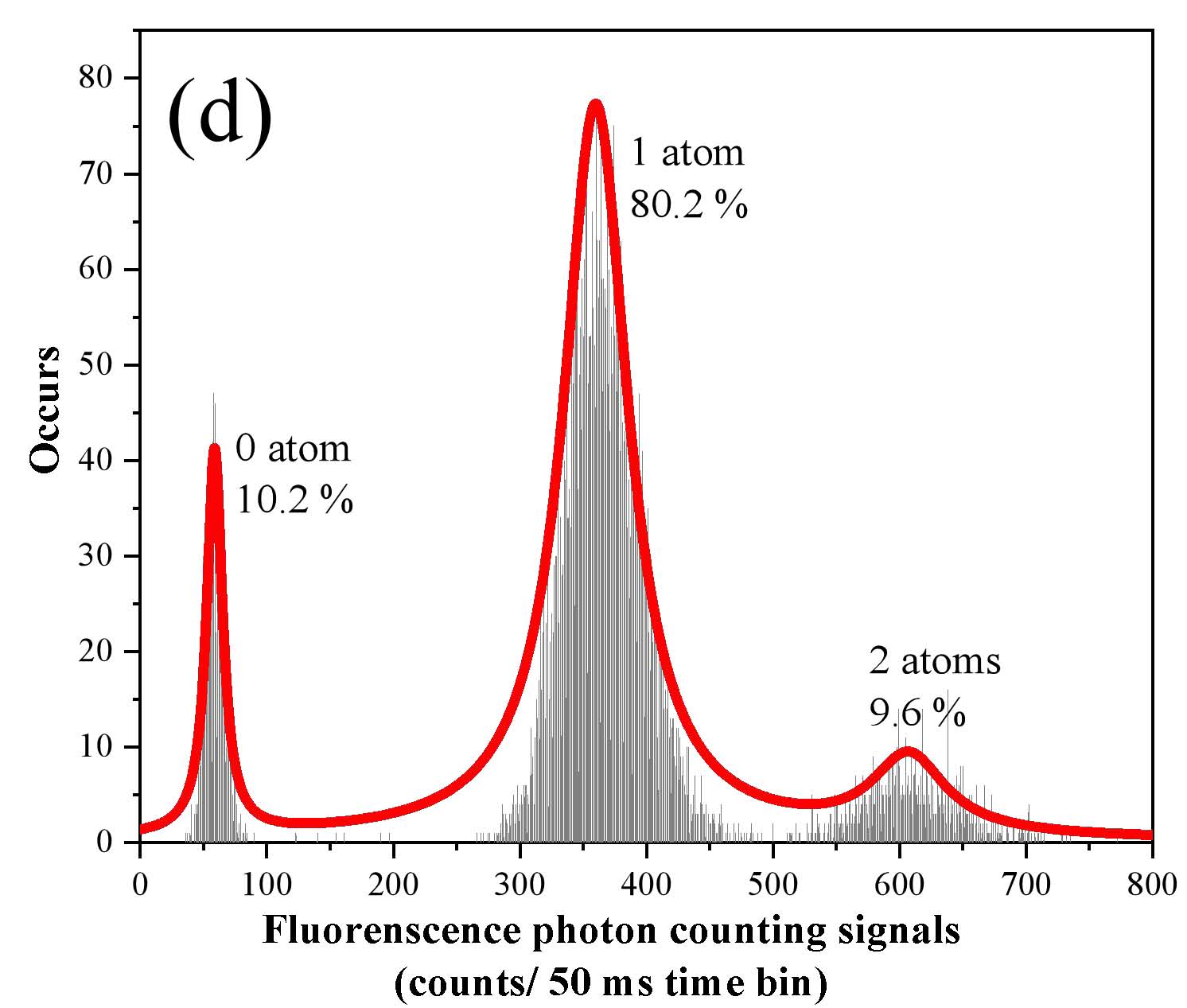}
                \label{fig:overview2_4}
        }
        \vspace{-0.15in}
        \caption{The loading probability of single atom in the MOT. The trigger loop can suppress the probability of multi-atom in the MOT and improve the probability of single atom. (a) and (b) When the trigger loop turned off, the loading probability of single atom in the MOT within 300 s is only 28.4\%, and the probability of multiple atom and zero atom is as high as 71.6\%, which is obviously not conducive to the measurement in subsequent experiments. (c) and (d) For the trigger loop on, the probability of single atom in the MOT increases to 80.2\%, while the probability of multi-atom loading is significantly suppressed.}
        \label{figure3}
\end{figure*}

Adjusting the spatial overlap and temporal overlap of the MOT and the optical tweezer can transfer the single atom between the traps efficiently. The loading rate of the MOT $R_L$ is sensitive to the axial gradient of the quadrupole magnetic field $(\frac{dB}{dZ})$, and $R_L$$\propto$$(\frac{dB}{dZ})$ [19,20]. The smaller the axial gradient of the quadrupole magnetic field, the higher the MOT loading rate and the more captured atoms. Conversely, the longer the magnetic field gradient, the lower the MOT loading rate and the fewer atoms trapped.

Since the loading rate of the MOT is pretty sensitive to the axial gradient of quadrupole magnetic field, we can use a trigger loop to control the loading rate of the MOT automatically. As shown in Figure 2(a), the fluorescence photons are collected into the single-photon counting module (SPCM), and the output pulses of SPCM enter the pulse counter. Then we can set the output voltage of the pulse counter for different circumstances that represents the atom numbers in the MOT. The output voltage of the pulse counter enters the quadrupole magnetic field power supply (Model SM 70-22, DELTA, the Netherland) as control voltage. The output current of the power supply will change related to the control voltage, and the quadrupole magnetic field gradient and the loading rate of the MOT change, too.

Figure \ref{figure2}$(b)$ and Table \ref{table:table1} show the working principle of the trigger loop. When there is only one atom in the MOT, the output voltage of the pulse counter keeps at 3.8 V, the quadrupole magnetic field gradient stays at 232 Gauss/cm. When there is no atom in the MOT, the output voltage of the pulse counter varies between 1.50 V and 1.79 V and yields the loading rate of the MOT going up. Under this circumstance, the output current of the power supply goes down automatically and the quadrupole magnetic field gradient change between 103 Gauss/cm and 116 Gauss/cm. For multi-atom condition, the output voltage of the pulse counter rises, and the loading rate of the MOT descends. On this occasion, the output current of the power supply rises and the quadrupole magnetic field gradient changes between 273 Gauss/cm and 316 Gauss/cm, and the atom number in the MOT cuts down.

Figure\ref{figure3} shows the probability of the single atom in the MOT in the cases of trigger loop off and on respectively. The power of the 852-nm MOT cooling laser is 150 $\mu$ W, and the power of the 894-nm repumping laser is 50 $\mu$W. When the trigger loop is off, the loading rate of the single atom is merely 28.4$\%$, and the two atoms loading rate reaches up to 56.8$\%$. The loading rate of single atom rises to 80.2$\%$ when the trigger loop is on, and the multi-atom rate declines sharply.

We can see in Figure\ref{figure3} that there is still certain photon-counting while there is no atom in the MOT. This is ude to the cooling and repumping laser beams hitting the glass cell's wall with scattering and reflecting, and the scattered photons at certain specific angles entering the fluorescence collection system. Moreover, the background atoms moving to the MOT laser path will interact with laser and emitted fluorescence photons will enter the fluorescence collection system, resulting in the generation of the background photon count. Besides, since the experimental conditions such as MOT cooling laser power and ambient temperature cannot be the same in each experiment, the background photon count in the MOT cannot be completely consistent but varies within a certain range.
\begin{figure*}[!htb]
        \centering
        \subfigure[] {
                \includegraphics[width=0.75\columnwidth]{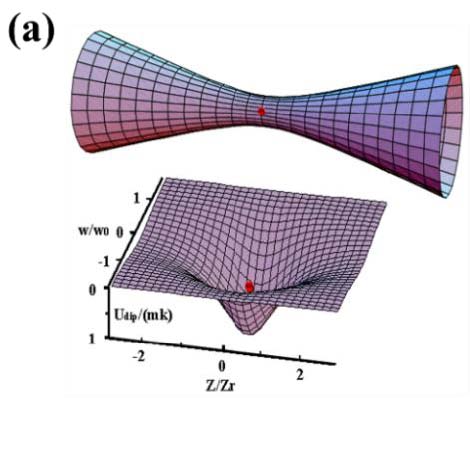}
                \label{fig:overview2_3}
            }
        \subfigure[]  {
                \includegraphics[width=0.9\columnwidth]{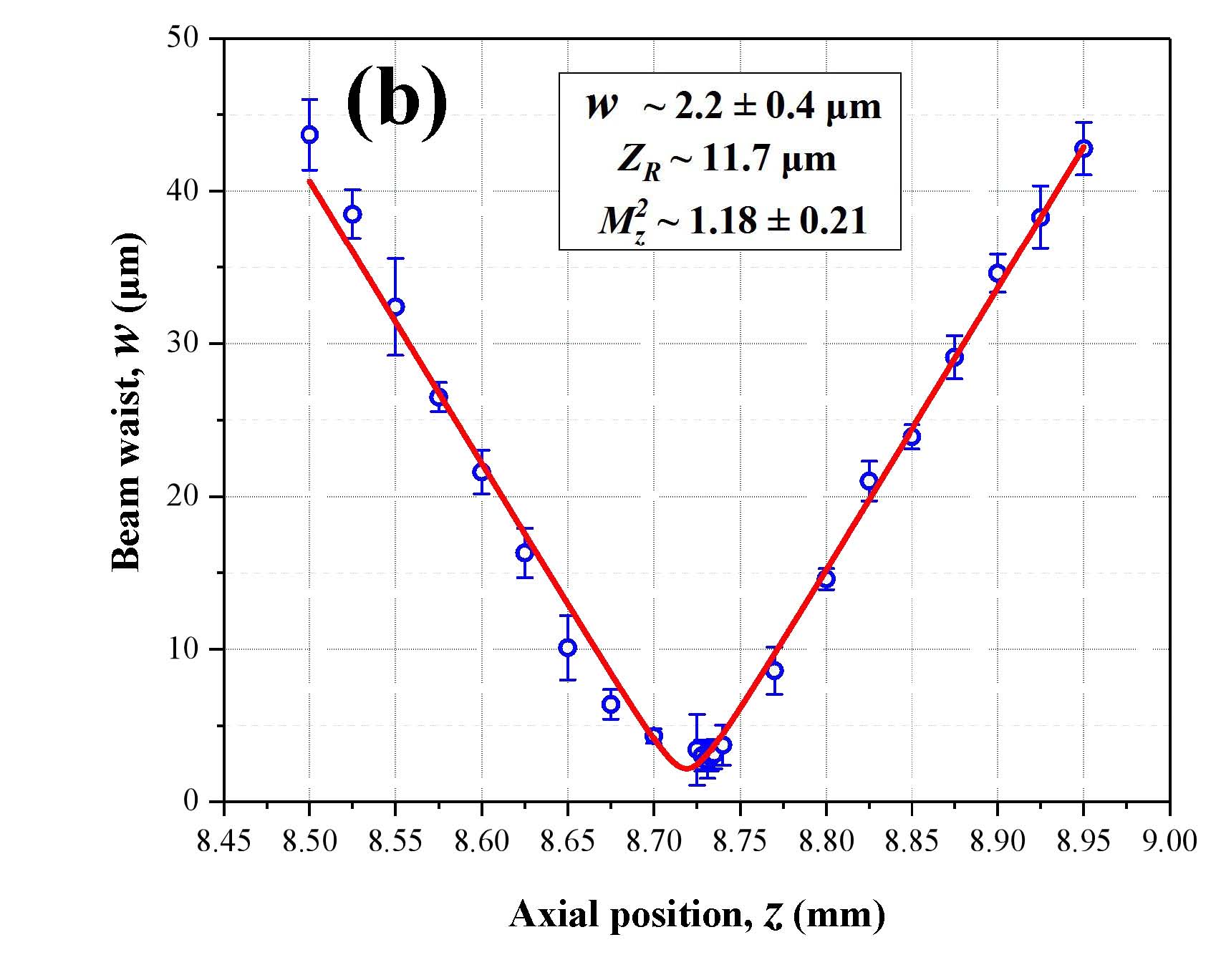}
                \label{fig:overview2_4}
        }
        \vspace{-0.15in}
        \caption{Scheme of optical tweezer. (a) A strongly focused single-spatial mode Gaussian beam of a red-detuned laser can confine atoms. (b) The beam waist radius and Rayleigh length of 1064-nm optical tweezer are 2.2 $\mu m$ and 11.7 $\mu m$, respectively..}
        \label{figure4}
\end{figure*}
\section{\label{sec:level1}Parametric Heating of Atoms in an Optical Tweezer}

Atoms in optical tweezers will generate electric dipole moments under the action of the gradient oscillating electric field. We can regard these atoms as electric dipoles interacting with the gradient oscillating external field. As shown in Figure\ref{figure4} (a) when the electric dipole is driven by an electric field with a frequency lower than the resonant frequency of atomic transition, due to the effect of attraction, the atom will be captured in the region with the strongest electric field. Otherwise, when the electric dipole is driven by an electric field with a frequency which higher than the resonant frequency, the atom will be excluded from the region with the strongest electric field. In other words, when the frequency of optical tweezer is red-detuning to the atomic transition frequency concerned, the atoms are captured in the optical tweezer\cite{21}. When the optical tweezer frequency is blue-detuning to the atomic transition frequency, the atoms will be excluded from the optical tweezer\cite{22}.

Using a focused single-spatial-mode Gaussian beam with red detuning can form an optical tweezer\cite{23}, with intensity distribution in the axial and radial direction as follows,
\begin{equation}
I(r,z)=\frac{2P}{\pi\omega^2(z)}\exp[\frac{-2r^2}{\omega^2(z)}]
\end{equation}
where P is the optical power of the optical tweezer laser beam, r is the radial component of the Gaussian beam, and w(z) is the beam's Gaussian radius at z \cite{23},
\begin{equation}
w(z)=w_0\sqrt{1+\frac{z^2}{z_R^2}}
\end{equation}
for $w_0$ is the beam's Gaussian radius, $z_R$ is the Rayleigh length, the axial and radial trap frequencies are 
$\omega_a=\sqrt{\frac{2U}{Mz_R^2}}$ and $\omega_r=\sqrt{\frac{4U}{w_0^2}}$ respectively, and U is the trap potential depth of the optical tweezer.

The Hamiltonian of single atom in the optical tweezer is \cite{23}
\begin{equation}
H=\frac{p^2}{2M}+\frac{1}{2}M\omega_{tr}^2[1+\varepsilon(t)]x^2
\end{equation}
for M is the mass of the Cs atom, $\omega_{tr}^2=k_0/M$ is the mean square of trap angular frequency,the elastic coefficient $k_0$ is in direct proportion to the light intensity of optical tweezer $I_0$, we can write the light intensity fluctuation as $\varepsilon(t)=\frac{I(t)I_0}{I_0}$. Here we use the first-order perturbation theory to clarify how the light intensity fluctuation acts on the single atom, and take the heating process as the transformation of the atom from $\vert n\rangle$ at $t=0$ to $\vert m\not= n\rangle$ at $t=T'$. The average transition probability $R_{m\leftarrow n}$ can express as the power spectral density of the laser intensity fluctuation $S_\varepsilon (\omega)$ \cite{23}
\begin{equation}
\int_0^\infty d\omega S_\varepsilon (\omega)=\int_0^\infty d\nu S_\varepsilon (\nu)=\langle \varepsilon^2(t)\rangle=\varepsilon_0^2
\end{equation}
for $\varepsilon_0$ is the mean square of intensity fluctuation and $\omega = 2\pi \nu$.

The average heating rate of atom in the optical tweezer is [23]
\begin{equation}
\langle \dot{E}\rangle=\sum_n P(n)2\hbar \omega_{tr}( R_{n+2\leftarrow n}- R_{n-2\leftarrow n})=\frac{1}{2}\omega_{tr}^2 S_\varepsilon (2\omega_{tr})\langle E\rangle
\end{equation}
where $P(n,t)$ is the probability of the atom stay at $\vert n\rangle$. For $\langle \dot{E}\rangle=\Gamma_\varepsilon \langle E\rangle$  the constant [23]
\begin{equation}
\Gamma_\varepsilon=\frac{1}{T_{I(sec)}}=\pi^2\nu_{tr}^2S_\varepsilon(2\nu_{tr})
\end{equation}

\begin{figure*}[!htb]
	\includegraphics[width=.8\textwidth]{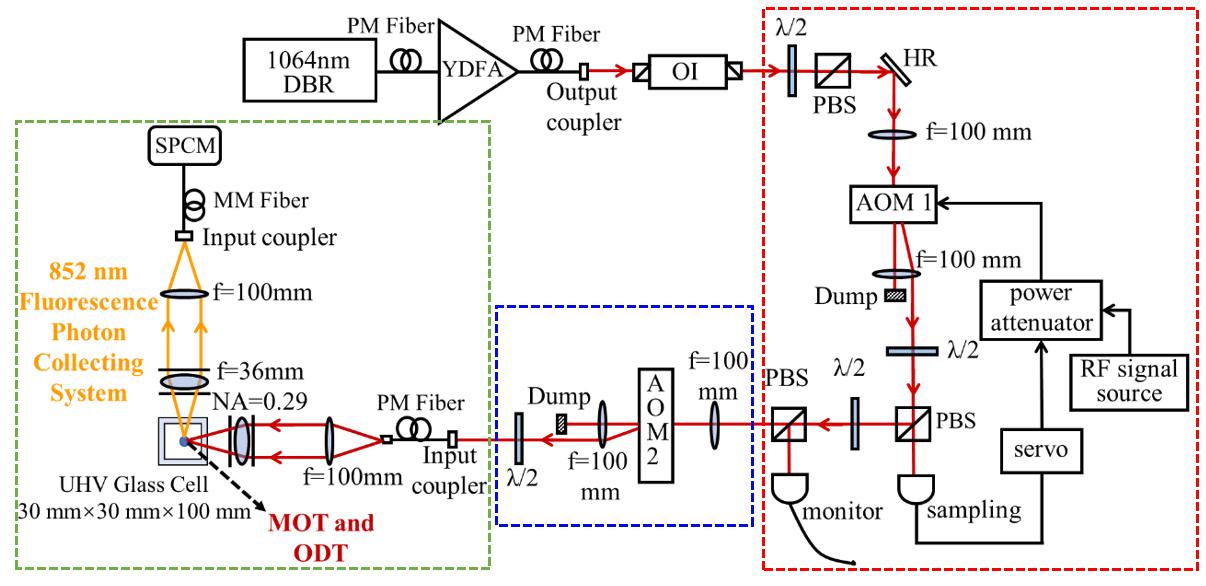}
	\caption{Experimental setup.  The red dash box shows the laser intensity fluctuation feedback loop by controlling the diffraction efficiency of AOM 1 to stabilize the intensity of the 1064-nm laser. The blue dash box shows the measurement setup of the trap frequency of optical tweezer. The green dash box shows the single atom cooling and trapping system, the anti-Helmholtz coils, but the cooling and repumping laser beams of the MOT are not shown here.Polarization-maintaining (PM) fiber; Multi-mode (MM) fiber; Magneto-optical trap (MOT); Optical-dipole trap (ODT), otherwise the optical tweezer; Optical isolator (OI); Acousto-optical modulator (AOM); Polarization beam splitter (PBS) cube; Single-photon counting module (SPCM); Ultra-high vacuum (UHV).}
	\label{figure5}
\end{figure*}

Among others, $\nu_{tr}$ is the trap angular frequency. According to Equation (8), the heating rate of single atom in the optical tweezer is related to the trap frequency and its double frequency of the optical tweezer. For a single atom in the optical tweezer, it will not only couple with the fluctuation resonance to the trap frequency of the tweezer, but also couple with the double-frequency more intensely. Ultimately, this heating mechanism makes the atom escaping from the optical tweezer. People call the process of converting the energy of the double frequency trap into the energy of the fundamental frequency trap as the parametric process. The heating effect of the intensity fluctuation of the laser on the single atom in the optical tweezer is the process of parametric heating.

\section{\label{sec:level1}Parametric Heating of Atoms in an Optical TweezeMeasurement of Trap Frequency of Optical Tweezer}

The output power of the 1064-nm laser (DBR-1064P, Thorlabs, USA) and Ytterbium-doped fiber amplifier (YDFA) system is up to 2 W. The 1064-nm laser beam passes a lens focal length of 100 mm to form a parallel beam with a diameter of 18–19 mm. Then the parallel beam goes through a lens assembly with a focal length of 36 mm and a numerical aperture of $NA=0.29$ to form the optical tweezers. Figure 4 (b) shows the measurement of the beam waist radius and $M_z^2$ factor of 1064 nm optical tweezer are 2.2 $\pm$ 0.4 $\mu$ m and 1.18 $\pm$ 0.21, respectively. The Rayleigh length $Z_R$ is 11.7 $\mu$ m.

As shown in Figure \ref{figure5}, the part in the blue box is used for measuring the trap frequency of the optical tweezer, which is an important parameter of the optical tweezer. We use a function generator (Model DG 635 SRS, USA) to apply modulation signals of different frequencies on AOM 2 (Model 3110-197, Crystal Technology, USA) to simulate the intensity fluctuation of the optical tweezer so that the optical tweezer will run with intensity fluctuation at a specific frequency.

Then, by measuring the transfer efficiency of single atom between the MOT and the optical tweezer to observe how the intensity fluctuation of different frequency acts on single atom trapping lifetime and Figure 6 shows a typical atomic transfer signal between the MOT and the optical tweezer. The transfer efficiency gets extremely low when the modulation frequency is resonant with the trap frequency and its double frequency of the optical tweezer.
\begin{figure}[!htb]
	\includegraphics[width=0.45\textwidth]{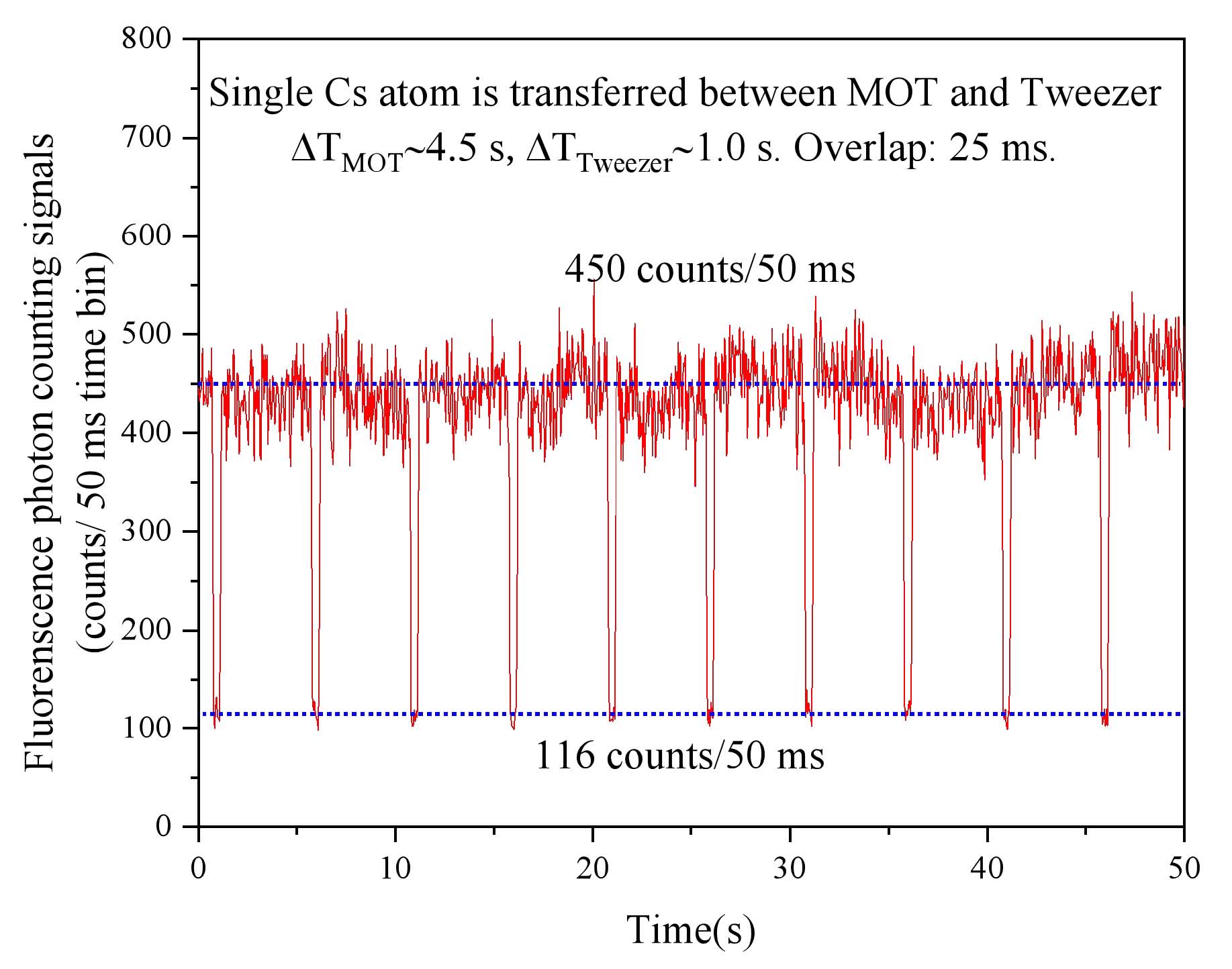}
	\caption{Atomic transfer signal between the MOT and the optical tweezer. The period is 5.5 s, in which the MOT duration is 4.5 s and the optical tweezer duration is 1.0 s. The MOT overlaps with the optical tweezer for 25 ms, the single atom is transferred between the MOT and the optical tweezer.}
	\label{figure6}
\end{figure}

\begin{figure}[!htb]
	\includegraphics[width=0.45\textwidth]{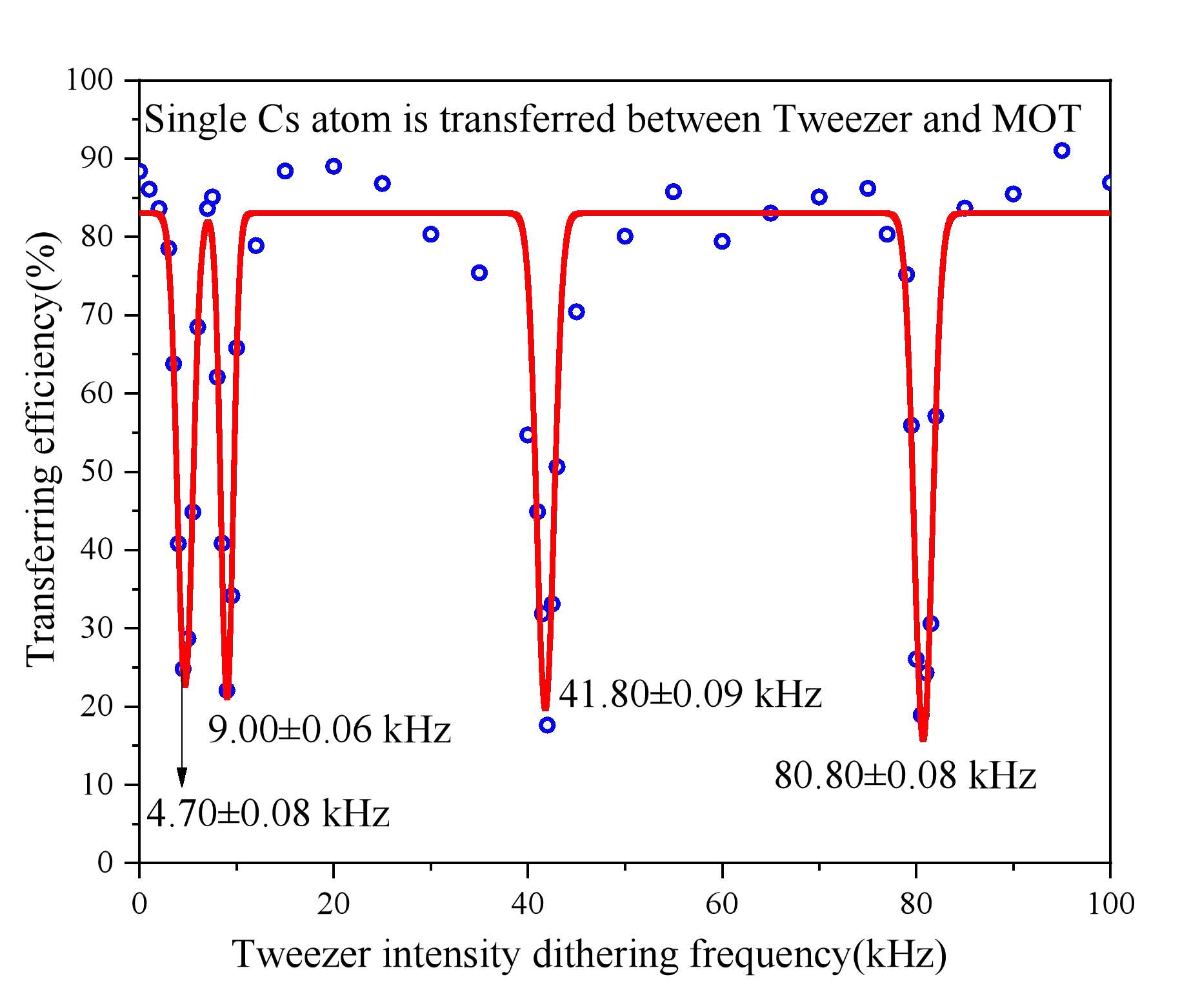}
	\caption{Trap frequency and double frequency of the 1064-nm optical tweezer. Transfer efficiency goes an apparent decline when the modulation frequency is resonant or near resonant to the trap frequency and its double frequency of the optical tweezer and reaches a minimum at resonance.}
	\label{figure7}
\end{figure}

Figure\ref{figure7} shows the trap frequency and its double frequency on the axial and radial direction. The power of 1064-nm optical tweezer is $\sim$48.5 mW, the optical tweezer trap's depth is $\sim$1.3 mK. The measured axial trap frequency is $\sim$4.70 kHz, for double frequency is $\sim$9.00 kHz (should be 9.40 kHz), its radial trap frequency is $\sim$41.80 kHz, and the double frequency is $\sim$80.80 kHz (should be 83.60 kHz).
The deviation between the experimental value and the theoretical calculation is mainly because the suppression effect of the laser intensity fluctuation feedback loop cannot be kept at the same level due to the long duration of the experimental process.

The photon scattering rate of the optical tweezer with trapped atoms inside is [24]
\begin{equation}
\Gamma_{SC}=\frac{3\pi c^2\omega_L^3}{2\hbar\omega_0^6}(\frac{\Gamma}{\omega_0^2-\omega_L^2}+\frac{\Gamma}{\omega_0^2+\omega_L^2})^2
\end{equation}
where $\Gamma$ is the spontaneous decay rate of the atomic transition, $\omega_0$ is the atomic transition frequency, and $\omega_L$ is the frequency of the optical tweezer. In a 1064-nm optical tweezer, Cs 6 $S_{1/2}$ $(F_g=4)$- 6 $P_{3/2}$ $(F_e=5)$ transition, the photon scattering rate is $\sim$8.3 photons/s.
\section{\label{sec:level1}Suppression of Laser Intensity Fluctuation}
\subsection{\label{sec:level2}Experimental Principles and Setup}

Figure\ref{figure5} shows the scheme of the laser intensity fluctuation control system. The part in the green box is the single atom capturing and observing system. The optical fiber's output beam goes through the collimating lens and expended to a near-parallel beam with a beam diameter of $\sim$19 mm. Then the parallel beam passing through the focusing lens assembly $(NA = 0.29)$, and becomes a tightly focused beam with a beam wrist of 2.2 $\mu$m Gaussian radius. We use the same lens assembly to collect the fluorescence photons of Cs atoms and steer them into the SPCM through a multi-mode (MM) fiber.

The part in the red box is the laser intensity fluctuation feedback loop. The first-order diffraction light of AOM 1 serves as the optical tweezer beam and the zero-order diffraction light is blocked by a dump. The first-order diffraction beam goes through the PBS, the reflected light is utilized for the sampling laser, collected into the detector (Model 2051, New Focus Inc., USA), and the laser intensity fluctuation can be charatorized by using of the detector’s output voltage. Then the output voltage signal enters the proportion integration differentiation (PID) (Model SIM960, SRS, USA) as the reference signal by which the feedback signal related parameters (e.g., integral time, etc.). Next, applying the feedback signal on the driving voltage of AOM 1, we can control the diffraction efficiency of AOM 1 precisely. When the optical tweezer laser (the first-order diffracted light of AOM 1) intensity decreases (or increases), the output voltage of the sampling detector decreases (or increases) too. The PID output positive feedback (or negative feedback) signal will make the diffraction efficiency of AOM 1increases (or decreases), which means the power of the first-order diffraction light goes up (or down). Finally, we can achieve the purpose of suppressing the fluctuation of laser intensity.

\begin{figure*}[!htb]
        \centering
        \subfigure[] {
                \includegraphics[width=0.9\columnwidth]{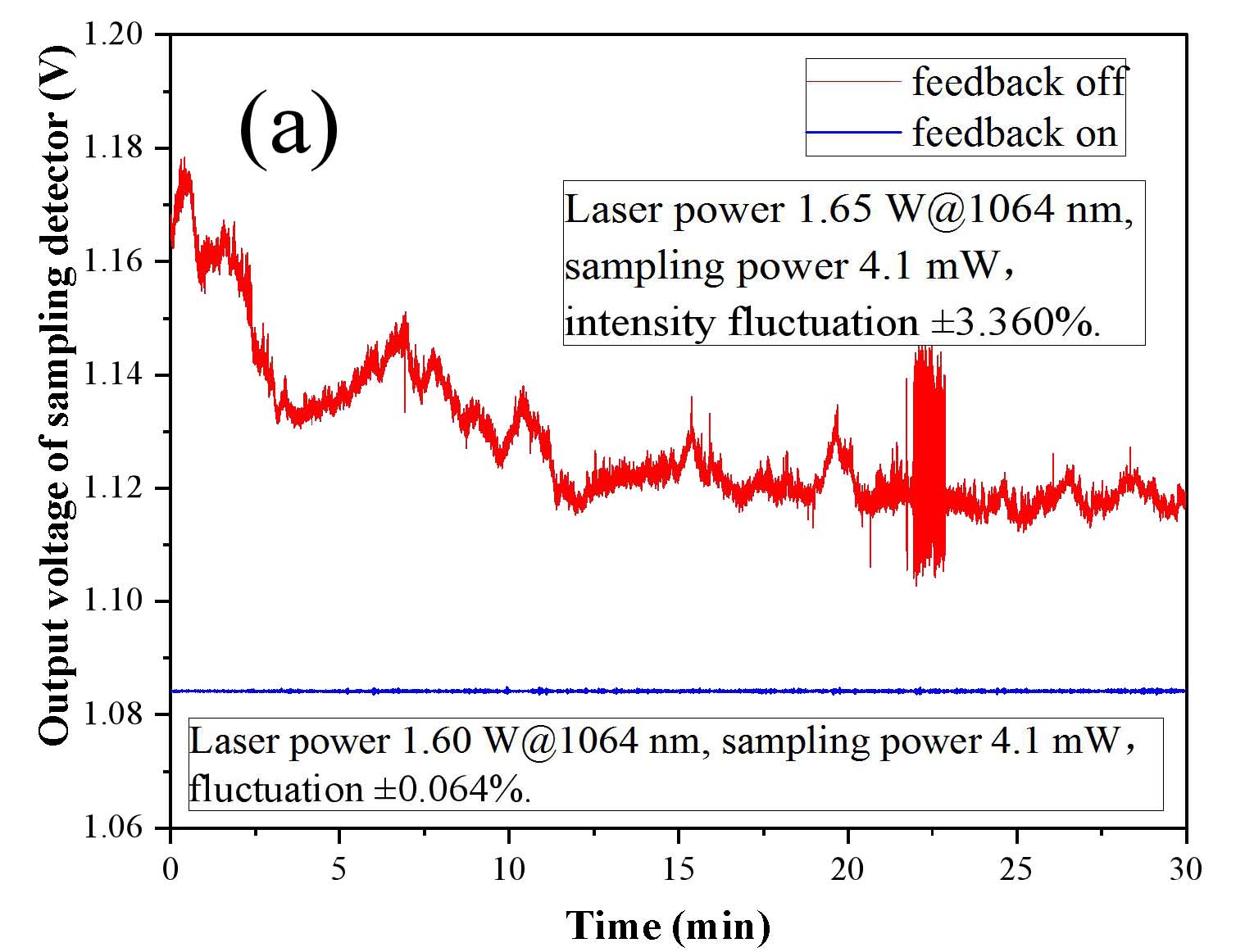}
                \label{fig:8a}
            }
        \subfigure[]  {
                \includegraphics[width=0.9\columnwidth]{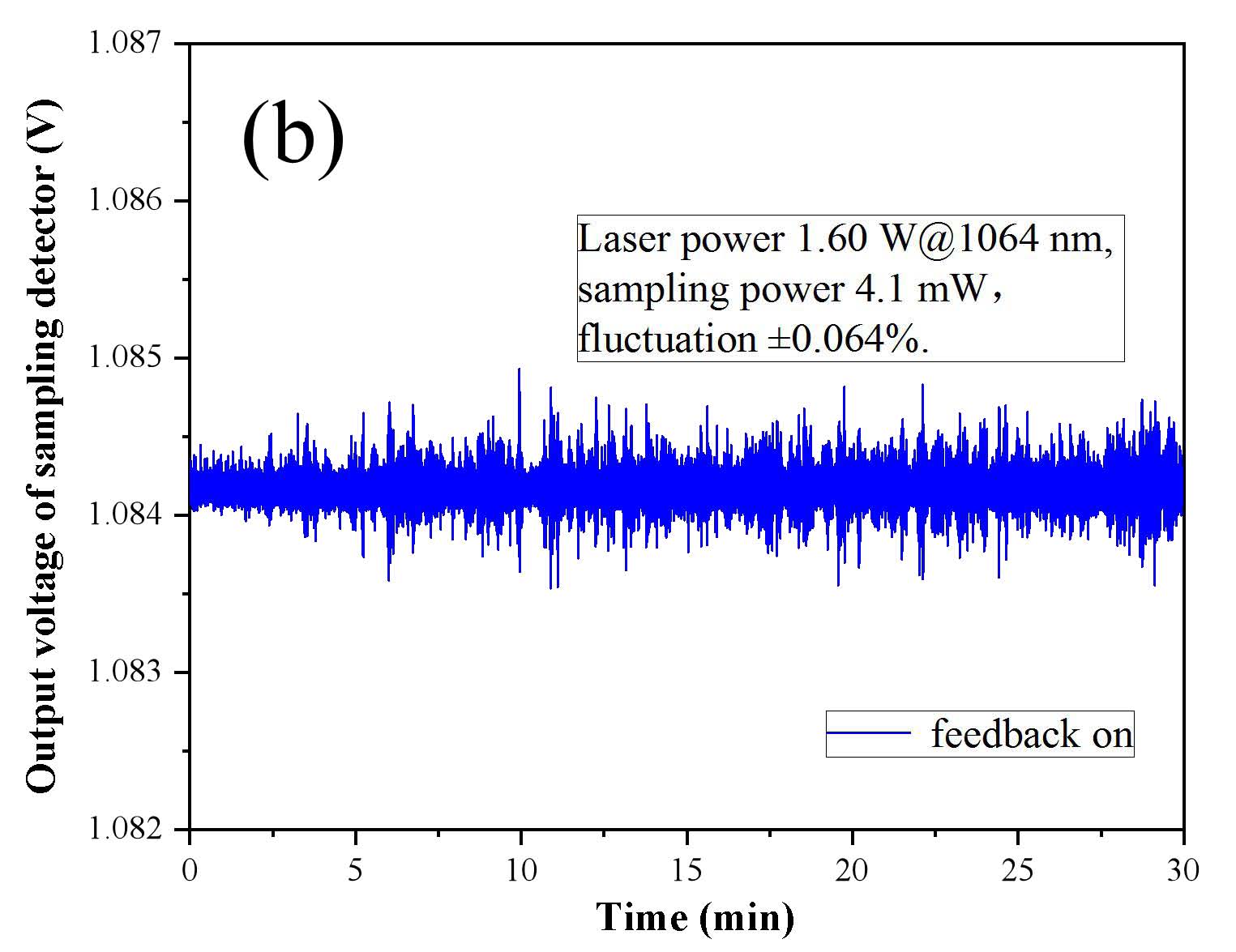}
                \label{fig:8b}
        }
        \vspace{-0.15in}
        \caption{Intensity fluctuation of laser beam in the time domain in the free-running case and the feedback-on case. The laser intensity fluctuation is suppressed from ±3.360\% to ±0.064\%.}
        \label{figure8}
\end{figure*}

\begin{figure}[!htb]
	\includegraphics[width=0.45\textwidth]{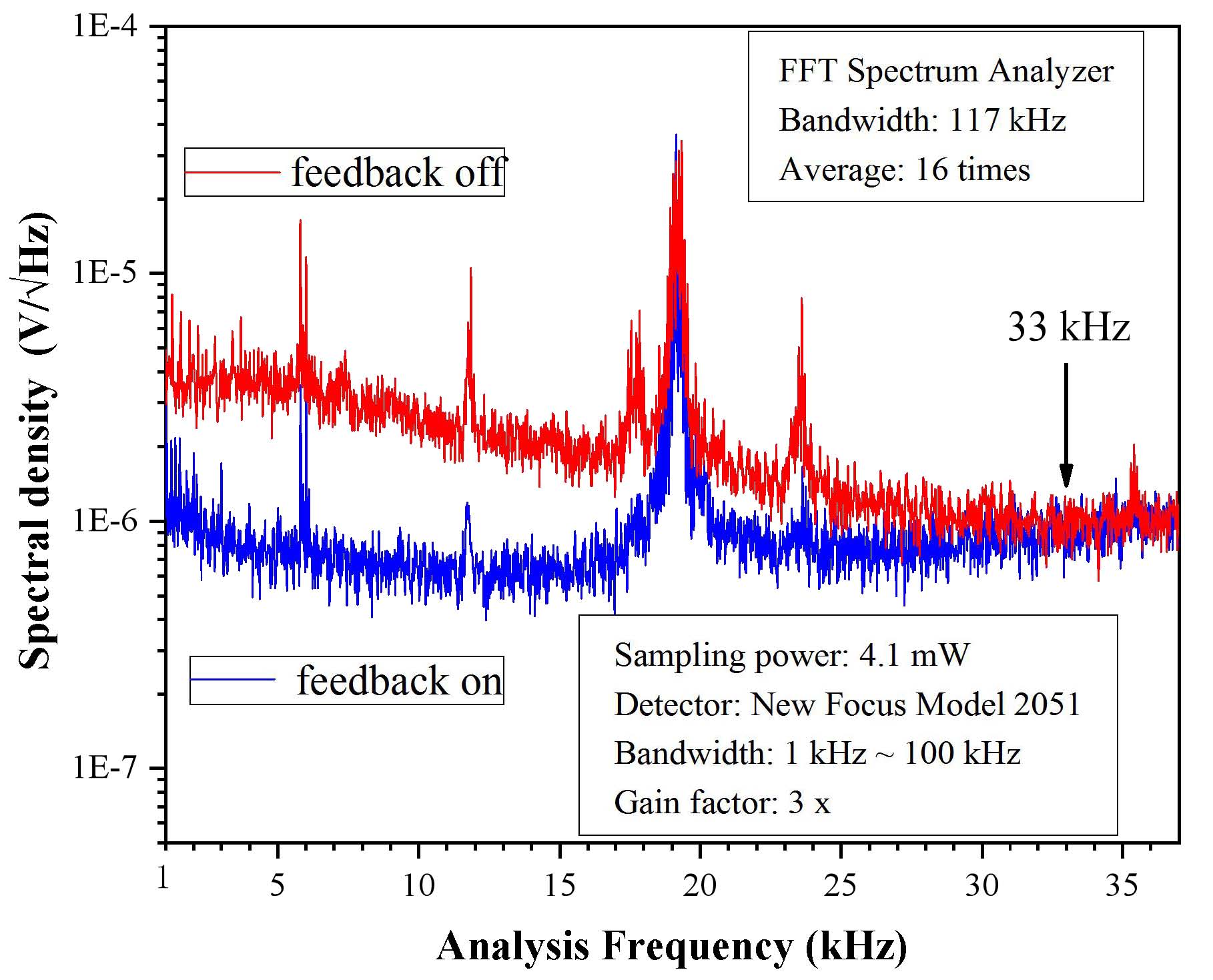}
	\caption{Comparison of the laser intensity fluctuation is in the frequency domain when feedback loop is on and off. The feedback bandwidth is $\sim$33 kHz. Several noise peaks came from the external environment and the YDFA’s cooling system.}
	\label{figure9}
\end{figure}

\subsection{\label{sec:level2}Suppression Results of Laser Intensity Fluctuation}

Figure\ref{figure8} and Figure\ref{figure9} show the response of the feedback loop. The total power of the 1064-nm laser beam is 1.6 W and the sampling power is 4.1 mW. The laser intensity fluctuation in the time domain for the free-running case and feedback-on case is $\mp$ 3.360$\%$ and $\mp$ 0.064$\%$, respectively. The feedback bandwidth $\sim$is 33 kHz, which covers the axial trap frequency of optical tweezer and its double frequency, and is much broader than that in our previous works [13,17]. There is a strong fluctuation peak on 19.1 kHz, due to the fluctuation from the external environment. The other four fluctuation peaks on 5.9 kHz and its double frequency (11.8 kHz), triple efficiency (17.7 kHz) and quadruple frequency (23.6 kHz) are caused by YDFA’s cooling system.

\section{\label{sec:level1}Improvement of Atom Trapping Lifetime in Optical Tweezers}
\begin{figure*}[!htb]
        \centering
        \subfigure[] {
                \includegraphics[width=0.815\columnwidth]{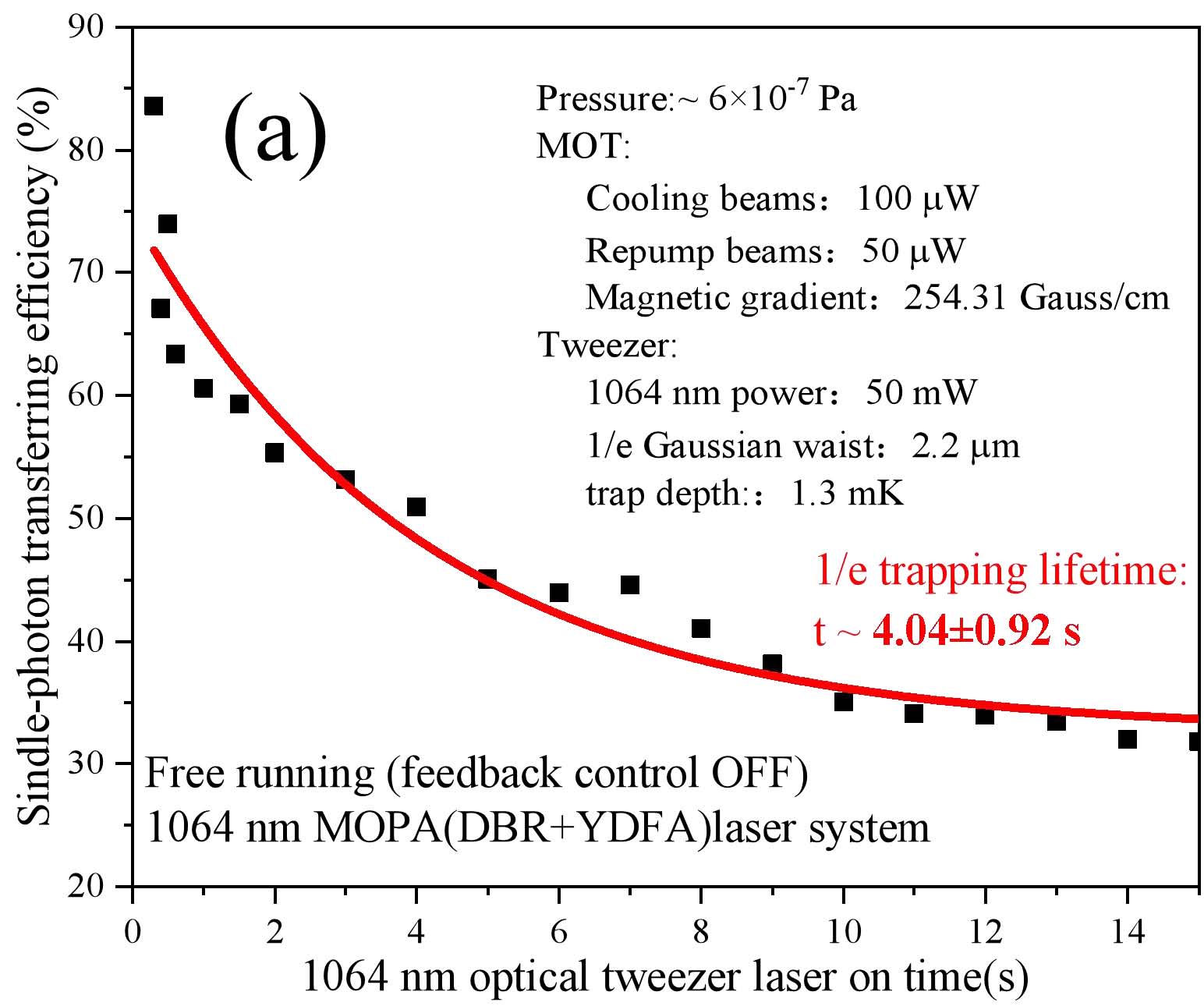}
                \label{fig:10a}
            }
        \subfigure[]  {
                \includegraphics[width=0.9\columnwidth]{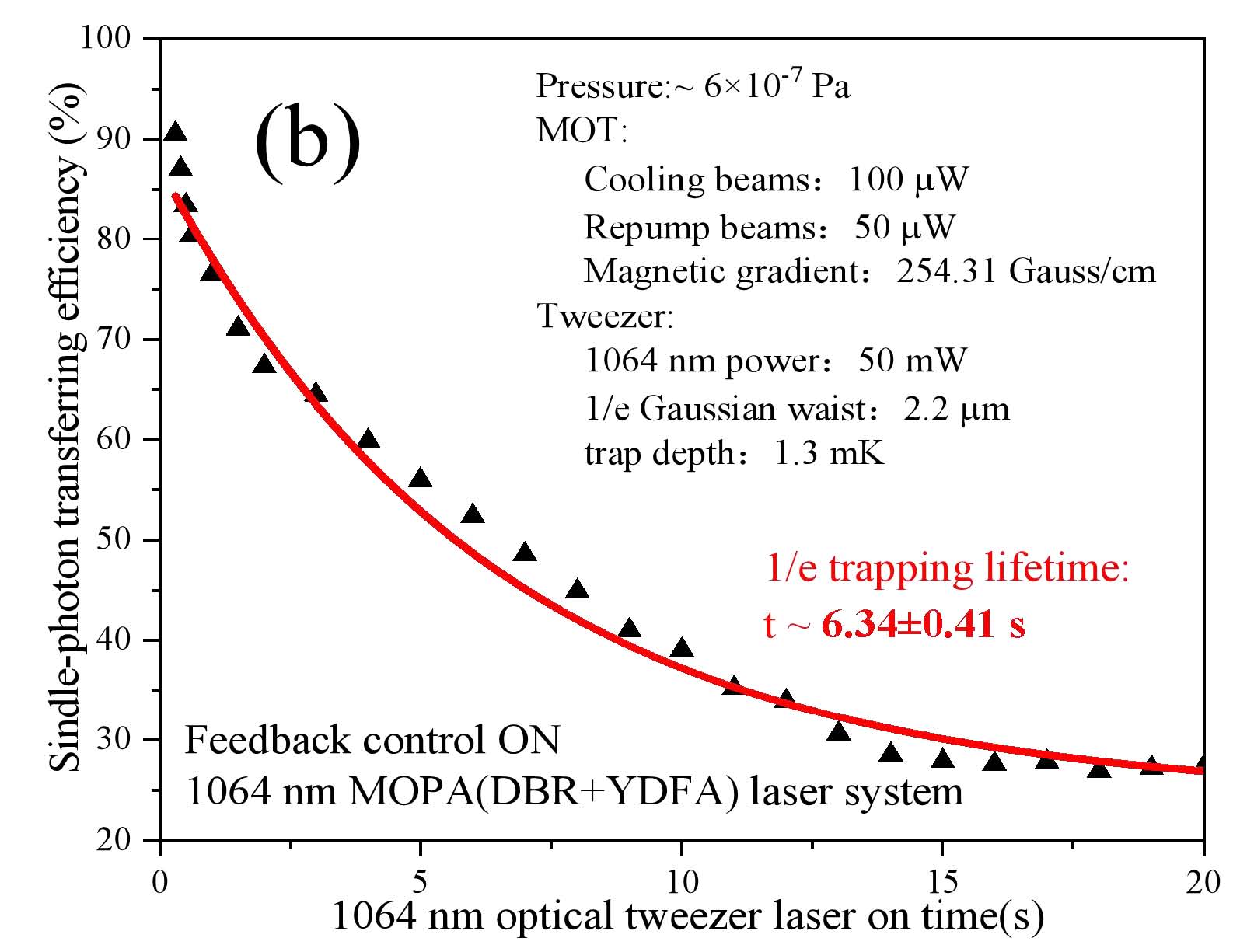}
                \label{fig:10b}
        }
        \vspace{-0.15in}
        \caption{Trapping lifetime of atom in 1064-nm optical tweezer when the feedback loop is turned off (a) and on (b). The trapping lifetime of single atom is extended from 4.04 s to 6.34 s.}
        \label{figure10}
\end{figure*}

In the experiment, the optical power of the 852-nm cooling laser is $\sim$105 $\mu$W, the optical power of the 894-nm repumping laser is $\sim$50 $\mu$W, the quadrupole magnetic field gradient along axial direction is 254 Gauss/cm, and the optical power of the 1064-nm optical tweezers is $\sim$50 mW. Firstly, we capture a single atom in the MOT and adjust the spatial overlap of the MOT and the optical tweezer. We change the experimental sequence to control the time overlap of the optical tweezer and the MOT for 25 ms, which maximizes the atom transfer efficiency between the MOT and the optical tweezer. Next, we measure the single atom transfer efficiency under different time duration of the optical tweezer. As the duration of optical tweezer increases, atom transfer efficiency goes down exponentially. Finally, the single atom trapping lifetime in the 1064-nm optical tweezer can be obtained by fitting measured data.

As shown in Figure\ref{figure10}, the trapping lifetime of single atom in the 1064-nm optical tweezer without the feedback loop on is 4.04 $\pm$ 0.92 s, while the trapping lifetime is 6.34 $pm$ 0.41 s when the feedback loop is turned on. We can see that the trapping lifetime of single atom in the optical tweezers is extended after the suppression of the laser intensity fluctuation of the optical tweezer.

The typical trapping lifetime here we achieved seems shorter than that of our previous work [19] because the background Cs density in the ultra-high vacuum (UHV) glass cell has increased a lot in this experiment. Now the trapping lifetime of single atom in the optical tweezer is mainly limited by the atomic collisions under the typical pressure of approximately 6 $\times$ $10^{-7}$ Pa.
\section{\label{sec:level1}Conclusions}

In this paper, we analyzed how the intensity fluctuation of optical tweezers act on the atomic trapping lifetime. We suppressed the intensity fluctuation of the 1064-nm optical tweezer to extend the trapping lifetime of single atom. The suppression bandwidth will be extended to cover the trap frequency and its double frequency on the both axial and radial direction. In addition, the suppression effect of laser intensity fluctuation and the feedback bandwidth can be adjusted to meet different experimental requirements, which would provide valuable insights for subsequent experiments of single atom manipulation and quantum simulation.


\noindent\textbf{Funding:} This research was funded by [The National Key R\& D Program of China] grant number [2017YFA0304502], [the National Natural Science Foundation of China] grant number [11774210, 6187511, 11974226, and 61905133], and [Outstanding Graduate Innovation Program of Shanxi Province] grant number [2019BY2016].


\begin{thebibliography}{999}
\bibitem{1}
Darquie, B.; Jones M.P.; Dingjan, J.; Beugnon, J.; Bergamini, S.; Sortais, Y.; Messin, G.; Browaeys, A.; Grangier, P. Controlled single-photon emission from a single-trapped two-level atom.,  {\em Science }, {\bf 2005}, {\em 309}, 454–456.


\bibitem{2}
Liu, B.; Jin, G.; Sun, R.; He, J.; Wang, J.M. Measurement of magic-wavelength optical dipole trap by using the laser-induced fluorescence spectra of trapped single cesium atoms., {\em Opt. Express}, {\bf 2017}, {\em 25}, 15861–15867.

\bibitem{3}
Liu, B.; Jin, G.; He, J.; Wang, J.M. Suppression of single-cesium-atom heating in a microscopic optical dipole trap for demonstration of an 852-nm triggered single-photon source., {\em Phys. Rev. A}, {\bf 2016}, {\em94}, 013409.


\bibitem{4}
Hänsch, T.W.; Schawlow, A.L. Cooling of gases by laser radiation., {\em Opt. Commun.}, {\bf 1975}, {\em 13}, 68–69.


\bibitem{5}
Raab, E.L.; Prentiss, M.; Cable, A.; Chu, S.; Pritchard, D.E. Trapping of neutral sodium atoms with radiation pressure., {\em  Phys. Rev. Lett.}, {\bf 1987}, {\em59}, 2631–2634.


\bibitem{6}
Hu, Z.; Kimble, H.J. Observation of a single atom in a magneto-optical trap.,{\em Opt. Lett.}, {\bf 1994}, {\em 19}, 1888–1890.


\bibitem{7}  
Barredo, D.; de Léséleuc, S.; Lienhard, V.; Lahaye, T.; Browaeys, A. An atom-by-atom assembler of defect-free arbitrary two-dimensional atomic arrays.,  {\em Science}, {\bf 2016}, {\em354}, 1021–1023.


\bibitem{8}  
Barredo, D.; Lienhard, V.; de Léséleuc, S.; Lahaye, T.; Browaeys, A. Synthetic three-dimensional atomic structures assembled atom by atom., {\em Nature}, {\bf  2018}, {\em 561}, 79–82.


\bibitem{9}  
He, J.; Yang, B.D.; Zhang, T.C.; Wang, J.M. Improvement of the signal-to-fluctuation ratio of laser-induced-fluorescence photon-counting signals of single atoms magneto-optical trap., {\em J. Phys. D: Appl. Phys.}, {\bf  2011}, {\em 44}, 135102.


\bibitem{10}  
He, J.; Wang, J.Y.; Yang, B.D.; Zhang, T.C.; Wang, J.M. single atoms transferring between a magneto-optical trap and a far-off-resonance optical dipole trap., {\em Chin. Phys. B}, {\bf 2009}, {\em 18}, 03404.


\bibitem{11}  
He, J.; Liu, B.; Diao, W.T.; Wang, J.Y.; Jin, G.; Wang, J.M. Efficient loading of a single neutral atom into an optical microscopic tweezer., {\em Chin. Phys. B}, {\bf 2015}, {\em 24}, 043701.


\bibitem{12}  
Jin, G.; Liu, B.; He, J.; Wang, J.M. High on/off ratio nanosecond laser pulses for a triggered single-photon source., {\em Appl. Phys. Express}, {\bf 2016}, {\em 9}, 072702.


\bibitem{13}  
Wang, J.C.; Sun, R.; Zhang, K.; Wang, X.; He, J.; Wang, J.M. Effect of single atom optical tweezer’s intensity fluctuation upon atom’s trapping lifetime and it’s improvement., {\em J. Quant. Opt.}, {\bf 2019}, {\em 25}, 180–186. (In Chinese)


\bibitem{14}  
He, J.; Yang, B.D.; Zhang, T.C.; Wang, J.M. Efficient extension of the trapping lifetime of single atoms in an optical tweezer by laser cooling., {\em Physica Scripta}, {\bf 2011}, {\em 84}, 025302.


\bibitem{15}  
Wen, X.; Han, Y.S.; Liu, J.Y.; He, J.; Wang, J.M. Polarization squeezing at the audio frequency band for the Rubidium D1 line., {\em Opt. Express}, {\bf 2017}, {\em 25}, 20737–20748.


\bibitem{16}  
Harb, C.C.; Ralph, T.C.; Huntington, E.H.; Freitag, I.; McClelland, D.E.; Bachor, H.A. Intensity-fluctuation properties of injection-locked lasers., {\em Phys. Rev. A}, {\bf 1996}, {\em 54}, 4370–4382.


\bibitem{17}  
Bai, L.L.; Wen, X.; Yang, Y.L.; Liu, J.Y.; He, J.; Wang, J.M. 397.5 nm ultra-violate laser stabilization based on feedback control via acousto-optic frequency shifter., {\em Chinese. J. Lasers}, {\bf 2018}, {\em 45}, 1001008. (In Chinese)


\bibitem{18}  
Monroe, C.; Swann, W.; Robinson, H.; Wieman, C. Very cold trapped atoms in a vapor cell. {\em Phys. Rev. Lett.}, {\bf 1990}, {\em 65}, 1571–1574.


\bibitem{19}  
He, J.; Yang, B.D.; Cheng, Y.J.; Zhang, T.C.; Wang, J.M. Extending the trapping lifetime of single atom in a microscopic far-off-resonance optical dipole trap., {\em Front. Phys.}, {\bf 2011}, {\em 6}, 262–270.


\bibitem{20}  
Haubrich, D.; Höpe, A.; Meschede, D. A simple model for optical capture of atoms in strong magnetic quadrupole fields., {\em Opt. Commun.}, {\bf 1993}, {\em 102}, 225–230.


\bibitem{21}  
Schlosser, N.; Reymond, G.; Protsenko, I.; Grangier, P. Sub-poissonian loading of single atoms in a microscopic dipole trap., {\em Nature}, {\bf 2001}, {\em 411}, 1024–1027.


\bibitem{22}  
Xu, P.; He, X.D.; Wang, J.; Zhan, M.S. Trapping a single atom in a blue detuned optical bottle beam trap., {\em Opt. Lett.},{\bf 2010}, {\em35}, 2164–2166.


\bibitem{23}  
Savard, T.A.; O’Hara, K.M.; Thomas, J.E. Laser-fluctuation-induced heating in far-off resonance optical traps., {\em Phys. Rev. A}, {\bf 1997}, {\em 56}, R1095–R1098.


\bibitem{24}  
Grimm, R.; Weidemüller, M.; Ovchinnikov, Y.B. Optical dipole traps for neutral atoms., {\em Adv. At. Mol. Opt. Phys.}, {\bf 2000}, {\em 42}, 95–170.


\end{thebibliography}
\end{document}